\PassOptionsToPackage{english}{babel}
\documentclass[%
 amsmath,amssymb,
 aps,
prb,
reprint,
floatfix,
]{revtex4-2}

\usepackage[final]{graphicx}
\usepackage{dcolumn}

\usepackage{braket}
\usepackage{array}

\newcommand{\dotkp}{\boldsymbol{k}\cdot\boldsymbol{p}}
\usepackage{url}

\begin{document}

\preprint{APS/123-QED}

\title{Efficient electronic structure calculations for extended systems of coupled quantum dots using a linear combination of quantum dot orbitals method}

\author{Alexander~Mittelst{\"a}dt}
 \email{mittelstaedt@tu-berlin.de}
\author{Ludwig~A.~Th.~Greif}%
\author{Stefan~T.~Jagsch}%
\author{Andrei~Schliwa}%
\affiliation{%
 Institut f{\"u}r Festk{\"o}rperphysik, Technische Universit{\"a}t Berlin, \\
 Hardenbergstr. 36, 10623 Berlin, Germany
}%

\date{\today}

\begin{abstract}
We present a novel \lq linear combination of atomic orbitals\rq --type of approximation, enabling accurate electronic structure calculations for systems of up to 20 or more electronically coupled quantum dots.
Using realistic single quantum dot wavefunctions as basis to expand the eigenstates of the heterostructure, our method shows excellent agreement with full 8-band $\dotkp$ calculations, exemplarily chosen for our benchmarking comparison, with an orders of magnitude reduction in computational time.
We show that, in order to correctly predict the electronic properties of such stacks of coupled quantum dots, it is necessary to consider the strain distribution in the whole heterostructure.
Edge effects determine the electronic structure for stacks of $\lesssim$\,10 QDs, after which a homogeneous confinement region develops in the center.
The overarching goal of our investigations is to design a stack of vertically coupled quantum dots with an intra-band staircase potential suitable as active material for a quantum-dot-based quantum cascade laser.
Following a parameter study in the In$_{x}$Ga$_{1-x}$As/GaAs material system, varying quantum dot size, material composition and inter-dot coupling strength, we show that an intra-band staircase potential of identical transitions can in principle be realized.
A species library we generated for over 800 unique quantum dots provides easy access to the basis functions required for different realizations of heterostructures.
In (the associated manuscript) Ref.\ [PRL], we investigate room temperature lasing of a terahertz quantum cascade laser based on a two-quantum-dot unit cell superlattice.
\end{abstract}

\maketitle

\section{Introduction}
The unique electronic properties of stacked quantum dots (QDs) offers enhancements not only for optical semiconductor devices such as lasers, optical amplifiers and single photon devices, but can also be advantageous for quantum cascade lasers (QCLs). In QCLs, amplification of radiation is realized via intra-band transitions of electrons running down a staircase potential generated by a semiconductor superlattice at an external bias. As proposed in Refs. \cite{suris_prospects_1996, dmitriev_quantum_2005, wingreen_quantum-dot_1997, chia-fu_hsu_intersubband_2000}, QD-based QCLs can benefit from intrinsically reduced electron-phonon scattering processes, free-carrier absorption processes and an intrinsically narrow gain spectrum of electronically coupled QDs resulting in improved temperature resilience and greatly reduced laser threshold current densities. The advantages of a three-dimensional confinement of carriers in QCLs are demonstrated by several theoretical studies and experiments on quantum cascade structures utilizing QDs \cite{dmitriev_quantum_2005, apalkov_influence_2003, vukmirovic_electron_2008, zhuo_quantum_2014}, but also using quantum well-based structures in strong magnetic fields splitting the 2D subbands into a series of Landau levels \cite{alton_magnetic_2003, tamosiunas_terahertz_2003, scalari_terahertz_2004, valmorra_ingaas_alingaas_2015}.\\
\begin{figure}[!t]
	\centering
	\includegraphics[width=2.4in]{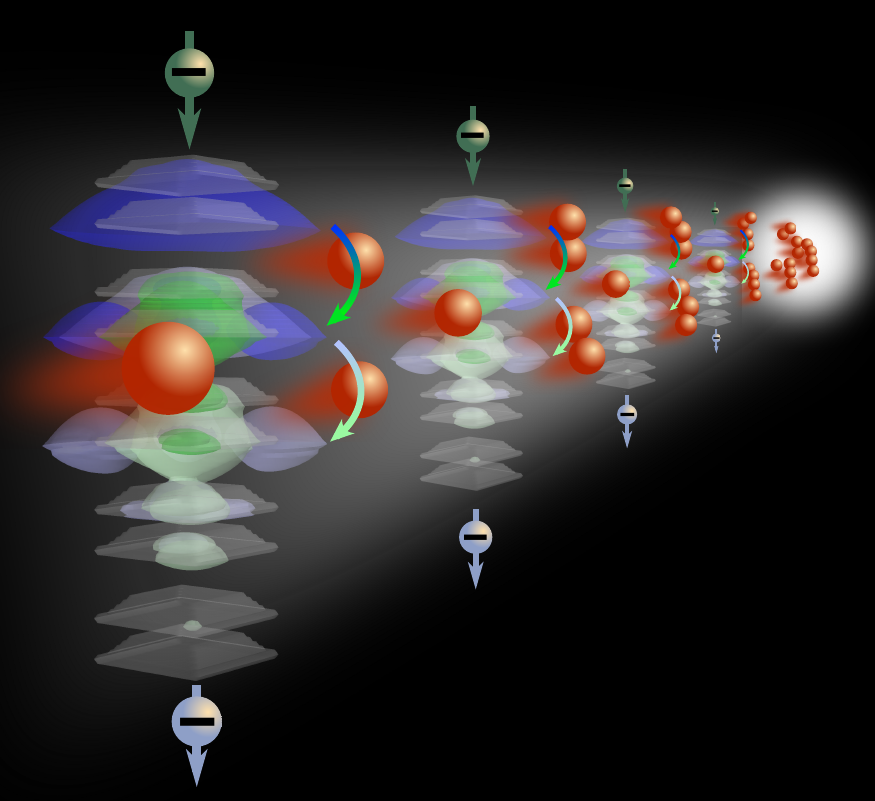} 
	\caption{Schematic illustration of light amplification in a QD cascade structure driven via an external bias. Stacks of electronically coupled QDs build a staircase potential providing optical transitions (blue electron densities) and states making non-radiative relaxation of carriers efficient (green electron densities).
	}
	\label{fig:fig_1}%
\end{figure}%
\begin{figure*}[!t]
	\centering
	\includegraphics[]{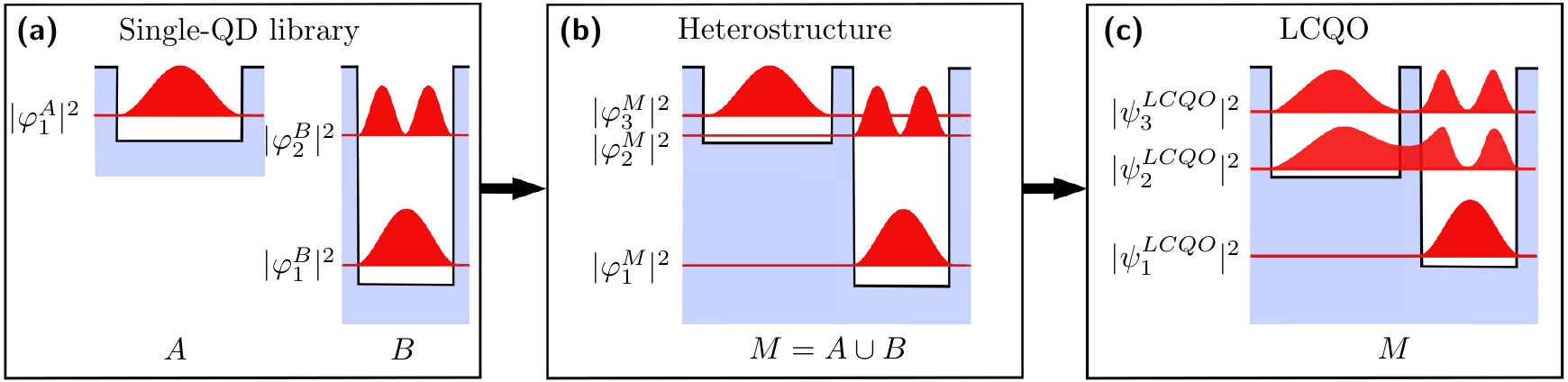} 
	\caption{
Schematic illustration of the LCQO method for system of two QDs. 
(a) A library of basis functions for different QD species ($A, B,\ldots$) is created. 
(b) The heterostructure is formed using the QDs from the library as building blocks ($M = A \cup B$).
The system's Hamiltonian is generated, considering the strain distribution and resulting piezoelectric potentials in the composite system $M$. 
(c) The eigenstates of the heterostructure $\ket{\psi_{i}^{LCQO}}$ are expanded in the union of the basis $\{\ket{\varphi_{k}^{M}}\} :=
\{\ket{\varphi_{j}^{A}}\} \cup \{\ket{\varphi_{l}^{B}}\}$.
    }
	\label{fig:fig_2}%
\end{figure*}%
A realization of a QCL based on a superlattice solely build of coupled QDs, cf. Fig.\ \ref{fig:fig_1}, however, is still lacking mainly for two reasons: 
On the one hand, a precise control over the desired structure during epitaxy, i.e. over the QD's size, shape, composition and their relative vertical position to each other, is needed. 
On the other hand, in order to arrive at a promising device design proposal, a rigorous simulation of the electronic structure of stacks of QDs requires a plenty of iterations already at the modeling stage.
The effort in finding suitable design parameters scales with the computational cost for an iteration of simulating the electronic structure of a realistic system of more than ten coupled QDs as well as calculating dozens of electronic states.
However, regarding the theoretical proposals and calculations of QD-based QCLs, the literature is based solely on assumptions of the electronic structure of such extended systems. Most investigations on the electronic properties of coupled QDs within the framework of tight-binding and 8-band $\dotkp$\,-models \cite{korkusinski_electronic_2001, jaskolski_strain_2006-1, usman_tight_2008, usman_experimental_2011, saito_optical_2008} involve just a stack of two QDs and a few involve nine layers of QDs calculating a maximum of twelve electronic states, where, in particular the tight-binding models are limited by large systems containing tens of millions of atoms.\\
In this paper, we demonstrate a novel \lq linear combination of atomic orbitals\rq --type of approximation making simulations of electronic properties of large systems of 20 and more electronically coupled QDs very efficient. 
Our \lq linear combination of quantum dot orbitals\rq --method (LCQO) works the following way:
Firstly, a library of QDs, systematically covering different sizes, shapes and chemical composition profiles, together with the associated single-particle eigenstates is created.
Then, driven by the desired target properties, systems of coupled QDs are virtually assembled together with the associated Hamiltonian, which is then expanded using the library of single-QD wave functions as basis.
The method as implemented herein uses basis functions calculated via an 8-band $\dotkp$\,-model including a realistic structure, strain and strain-induced internal fields, but can be implemented as such on any atomistic or continuum model. 
The resulting LCQO-eigenstates show an excellent agreement in a side-by-side comparison with a full 8-band $\dotkp$\,-simulation for stacks of two QDs and a benchmark comparison of the two methods for exemplarily chosen system of 10 stacked QDs shows a reduction of at least three orders of magnitude in computational time.
By using the LCQO approach, we are able to take into account the individual three-dimensional morphology of each QD within the stack, calculate strain and strain-induced internal fields of the whole structure as well as inter-dot electronic coupling.\\
We show why simulating 10 or more QDs is necessary to obtain uniform conduction band energy splittings and electron probability densities along the QD-chain, highlighting the influence of strain and piezoelectric fields on the bandstructure of electronically coupled QDs. 
A simulation of  $~$\,20 QDs is therefore necessary to get an intra-band staircase potential of several identical transitions as adaptable for a QD cascade active region design.\\
This paper is organized as follows, after introducing the method of calculation and a comparison to an 8-band $\dotkp$\,-simulation,
we analyze the electronic properties of stacks of In$_{x}$Ga$_{1-x}$As QDs embedded in a GaAs matrix with a focus on conduction band intra-band transitions as applicable in QCLs based on a QD superlattice and demonstrate the necessity of simulating the whole heterostructure.
Subsequently, we provide a parameter study by varying quantum dot size, material composition and inter-dot coupling strength of stacks of QDs. 
The last part shows that an uniform staircase potential of identical transitions of an exemplary QD heterostructure can be realized and highlights the impact of applying an external bias. 
Based on our library of more than 800 unique QDs making the LCQO approach very efficient in finding QD species and inter-QD spacings suitable for designing a QCL based on a QD superlattice, we developed in (the associated manuscript) Ref.\ [PRL] an active region design providing optical transitions within the terahertz regime. 
This active region comprises stacks of QD providing a two-QD unit cell superlattice, where we investigate room temperature lasing of a corresponding QCL device\\
\section{The linear combination of quantum dot orbitals method}
\begin{figure}[!t]
	\centering
	\includegraphics[]{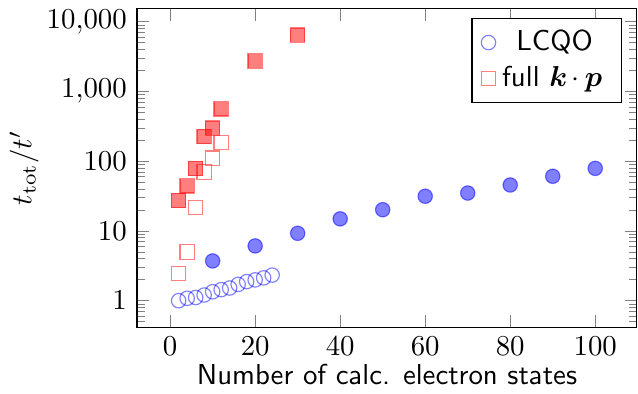}
	\caption{A comparison of calculation time between the 8-band $\dotkp$-simulations (red squares) and the LCQO method (blue circles). Empty and filled symbols denote a stack of two and ten QDs, respectively. The total calculation time $t_{\textrm{tot}} $ is normalized to the time of an LCQO simulation calculating two states in stack of two QDs.
	The simulations were performed on AMD Opteron 6274 processors, where the $\dotkp$-simulations were parallelized using four threads.
	}
	\label{fig:fig_3}%
\end{figure}

Along the lines of a linear combination of atomic orbitals (LCAO) method, a large system $M$ of coupled QDs is split into subsystems $I$ of single QDs, so that
\begin{align}
M := \bigcup \{I\mid I \ \textrm{is a single QD subsystem}\},
\end{align} 
for which sets of single-particle states $\lbrace\ket{\varphi_{1}^{I}}, \dots,\ket{\varphi_{n}^{I}}\rbrace$ can be efficiently calculated.
As illustrated in Figs.\ \ref{fig:fig_2}(a) and (b), these single-particle wavefunctions are calculated for different QD species building a library of wavefunctions then used as basis in the LCQO approximation of the eigenstates of the composite system
\begin{align}
\label{eq:expansion1}
\ket{\psi_{i}^{LCQO}} = \sum_{k=1}^{m} a_{ik} \ket{\varphi_{k}^{M}}\,,
\end{align}
where $\{\ket{\varphi_{k}^{M}}\} := \cup \{\ket{\varphi_{j}^{I}}\}$.
In this way, the number of coefficients $\{a_{1}(\boldsymbol r),\ldots,a_{m}(\boldsymbol r) \}$ in the variational problem of finding the eigenstates of $M$ is reduced to $m=|M|\times n$, where $|M|$ is the number of coupled QDs and $n$ is the number of single-particle states used as basis functions per QD.
The number of coefficients $m$ is therefore independent of the number of grid points of the $\ket{\varphi_{k}^{M}}$.
We note that the amount of basis functions $n$ for each QD is not necessarily a constant and could, for example, be the number of bound states for each QD, in a system composed of QDs differing in size or material composition.
When creating the basis for the LCQO method, the goal is to approximate the eigenstates of the heterostructure as best as possible with a limited set of basis functions, which is achieved by using realistic single QD wavefunctions.
In order to find the eigenstates of $M$, we adopt the Rayleigh-Ritz variational principle for the energy functional \cite{evarestov_quantum_2007}. 
Using the expansion in Eq.\ \ref{eq:expansion1},
\begin{align}
\varepsilon_{i}[\psi_{i}^{LCQO}] &= \frac{\braket{\psi_{i}^{LCQO}|\boldsymbol{H}|\psi_{i}^{LCQO}}}{\braket{\psi_{i}^{LCQO}|\psi_{i}^{LCQO}}} \\
\label{eq:energy_functional}
&= \sum_{k,l=1}^{m}\frac{ a_{il}^* a_{ik} \overbrace{\Braket{\varphi_{l}^{M}|\boldsymbol{H}|\varphi_{k}^{M}}}^{H_{lk}}}{ a_{il}^* a_{ik}\underbrace{\Braket{\varphi_{l}^{M} | \varphi_{k}^{M}}}_{S_{lk}}}\, ,
\end{align}
where $\boldsymbol{H}$ denotes the Hamiltonian of the composite system.
Notably, the system Hamiltonian $\boldsymbol{H}$ contains not a mere superposition of the electrostatic potentials of the QD subsystems, which would be an oversimplification for systems of QDs, cf.\ section III.B, but instead considers the strain distribution in $M$. 
Since the strain state and the resulting piezoelectric fields in the QD heterostructure strongly depend on the actual QD configuration as well as geometry and material composition of the individual QDs, the Hamiltonian has to be generated individually for each unique assembly.
Varying the energy functional in Eq.\ \ref{eq:energy_functional} with respect to the expansion coefficients $\delta \varepsilon_{i} / \delta a_{il}^*$, and minimizing the energy, leads to the generalized eigenvalue problem 
\begin{align}
\label{eq:ci}
\left(\begin{matrix}
H_{11} & \cdots & H_{1m} \\
\colon & \ddots & \colon\\
H_{m1} & \cdots & H_{mm} \\
\end{matrix}\right)
\left(\begin{matrix}
a_1 \\
\colon\\
a_m\\
\end{matrix}\right) = \varepsilon
\left(\begin{matrix}
S_{11} & \cdots & S_{1m} \\
\colon & \ddots & \colon\\
S_{m1} & \cdots & S_{mm} \\
\end{matrix}\right)
\left(\begin{matrix}
a_1 \\
\colon\\
a_m\\
\end{matrix}\right)
\end{align} 
that yields $m$ eigenvalues $\varepsilon_{i}$ and corresponding eigenvectors $\ket{a_{i}}$, containing the coefficients $a_{ik}$ of the LCQO eigenfunctions $\ket{\psi_{i}^{LCQO}}$, cf. Fig.\ \ref{fig:fig_2}(c).\\
So far the LCQO method is universal and could be implemented on top of any atomistic or continuum model used to calculate the electronic states of the single QDs.
In the present work, we exemplarily use an established 8-band $\dotkp$ model, including strain-induced internal fields up to second-order piezoelectricity, to calculate the single-particle wavefunctions of the individual QDs \cite{grundmann_inas_gaas_1995, stier_electronic_1999, schliwa_impact_2007}.
Within the 8-band $\dotkp$ model, the LCQO basis is expanded according to
\begin{align}
\label{eq:expansion2}
\ket{\varphi_{k}^{M}} = \sum_{o=1}^{8} b_{ko} \ket{\xi_{o}}, 
\end{align}
with the complex-valued envelope-function coefficients $b_{ko}$ and the atom-like Bloch functions $\ket{\xi_{o}}$. 
The LCQO-eigenfunctions in Eq. \ref{eq:expansion1} are then given as
\begin{align}\label{eq:expansion3}
\ket{\psi_{i}^{LCQO}} = \sum_{k=1}^{m} a_{ik}' \sum_{o=1}^{8} b_{ko} \ket{\xi_{o}}.
\end{align}
A performance benchmarking of the LCQO method in comparison to a full 8-band $\dotkp$ calculation for exemplary systems of 2 and 10 QDs is provided in Fig.\ \ref{fig:fig_3}.
The benchmark reveals already for the calculation of 20 single-particle states in a stack of 10 QDs a reduction of at least three orders of magnitude in computational time for the LCQO method compared to a full 8-band $\dotkp$-calculation. Since the calculation time for the $\dotkp$ method increases strongly with the number of states, it is impractical to calculate more states, while with the LCQO method 50 and more eigenstates can be calculated very effectively.
\begin{figure}[!t]
	\centering
	\includegraphics[]{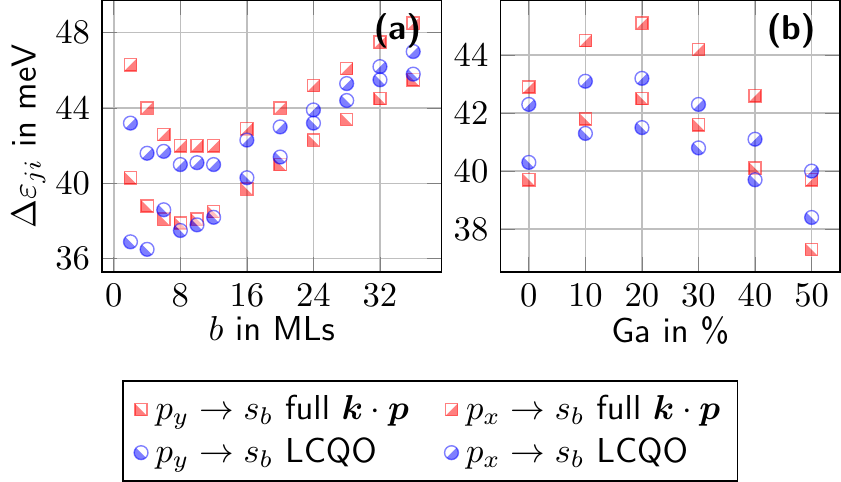} 
	\caption{Evolution of the energetic positions of the $p$-to-$s$ transitions for a stack of two identical QDs as a function of barrier width (a) and material composition (b) for the $\dotkp$\,-model (red squares) and the LCQO (blue circles). Semifilled circles and squares distinguish the $p_y$- and $p_x$-to-$s$ transitions.}
	\label{fig:fig_4}%
\end{figure}%
\section{Results} \label{ch:results}
\begin{figure*}[!t]
	\centering
	\includegraphics[]{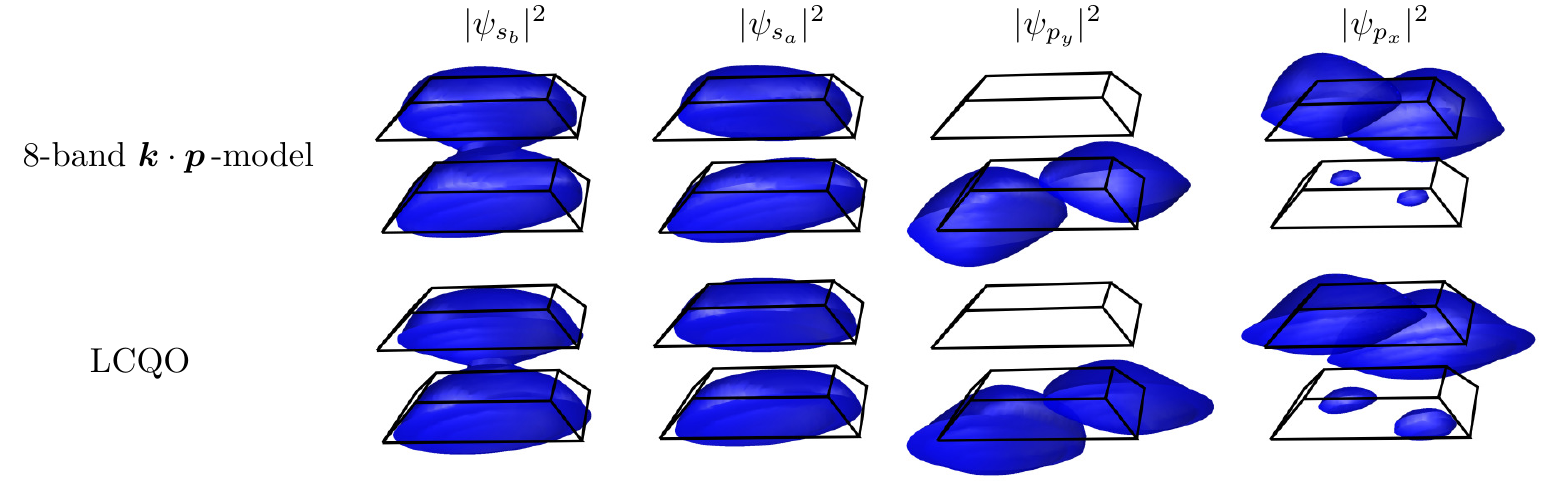}
	\caption{Probability densities of the first four Kramers-degenerated electron states (isosurface at 90\%) for the 8-band $\dotkp$\,-model and the LCQO method, respectively. The system considered is a stack of two identical In$_{1.0}$Ga$_{0.0}$As QDs with a barrier width of $b=16\,$MLs showing binding and anti-binding $s$-type orbitals, $|\psi_{s_b}|^2$ and $|\psi_{s_a}|^2$ respectively, followed by $p$-type orbitals, $|\psi_{p_y}|^2$ and $|\psi_{p_x}|^2$, respectively.
	}	
	\label{fig:fig_5}%
\end{figure*}%
We investigate the electronic structure of vertically coupled In$_{x}$Ga$_{1-x}$As/GaAs QDs, with the goal to design a heterostructure with an intra-band staircase potential suitable for a QD-QCL active region.
A decisive advantage of the LCQO method is that a library of basis functions for realistic single QDs can be prepared to facilitate parameter series of coupled QD systems, cf.\ Fig.\ \ref{fig:fig_2}.
We created such a species library for over 800 In$_{x}$Ga$_{1-x}$As/GaAs QDs varying in size, vertical aspect ratio $AR_{v}$ (height $h$ divided by the base diameter $d_{b}$) and material composition, within experimentally realistic limits.
In agreement with TEM investigations in Refs.\ \cite{blank_quantification_2009, litvinov_influence_2008}, the In$_{x}$Ga$_{1-x}$As QDs are modelled as truncated pyramids with a side-wall inclination of $40^{\circ}$, embedded in a GaAs matrix. 
With the exception of the parameter series in section III.C, results are presented for a model QD with a base diameter of $20.8\,$nm and a height of $2.8\,$nm ($AR_{v}=0.135$), along experimental reports in Refs.\ \cite{yamauchi_electronic_2006, bruls_stacked_2003, sugaya_multi-stacked_2011,  lemaitre_composition_2004}. In comparison to the literature, our model QD has a slightly smaller $AR_{v}$ to account for material inter-diffusion in real systems, since experiments show QDs exhibiting a composition gradient with a decreasing In-content, which in turn results in a confinement region smaller than the QDs geometry.
As basis functions per QD we calculate the five lowest Kramers-degenerate electron states (10 states per QD), which is the number of bound electrons in the model QD. This base size results in an excellent agreement of the energy eigenvalues and envelopes, as shown below for the states in a stack of two QDs. In addition, the Supplemental Material at [URL] shows that already with a base of $n=10$ the energy eigenvalues $\varepsilon_i$ of the lowest four Kramers-degenerated orbitals converge against the values of the full $\dotkp$ calculation within a $0.1$\% error.
To solve the eigenvalue problem in Eq.\ \ref{eq:ci}, we use a standard linear algebra package (\textsc{lapack}) with a finite difference grid resolution of two monolayers (MLs) of GaAs ($5.653\,$\AA), equally in case of the full $\dotkp$-calculations.
\subsection{A direct comparison between LCQO and a full 8-band $\dotkp$ calculation} \label{cp:validity}
\begin{figure*}[!t]
	\centering
	\includegraphics[]{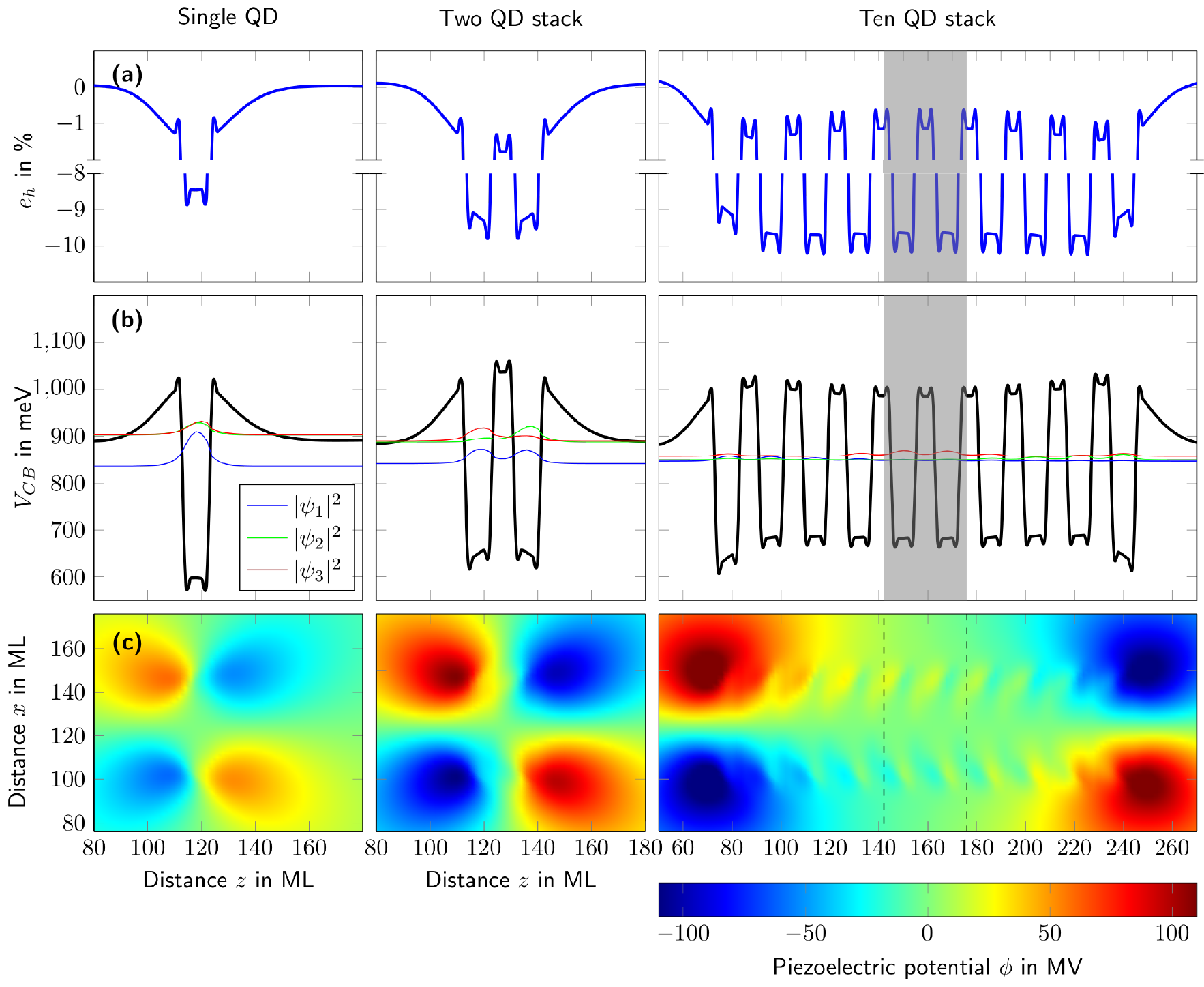}
	\caption{
	Diagrams illustrating the hydrostatic strain, piezoelectricity and the conduction band edge for a single QD and stacks of two and ten identical In$_{1.0}$Ga$_{0.0}$As QDs. Here, the barrier width is set to $b=8\,$MLs and the wetting layer is omitted. (a) The distribution of the hydrostatic strain $e_{h}$ and (b) the evolution of the conduction band edge $V_{CB}$. (c) The corresponding distribution of the piezoelectric potential within the $(001)$-plane trough the center of the QD stacks. The $|\psi_{i}(z)|^{2}$ show the probability densities of the first three Kramers-degenerate electron states.}
	\label{fig:fig_6}%
\end{figure*}%
\begin{figure}[!t]
	\centering
	\includegraphics[]{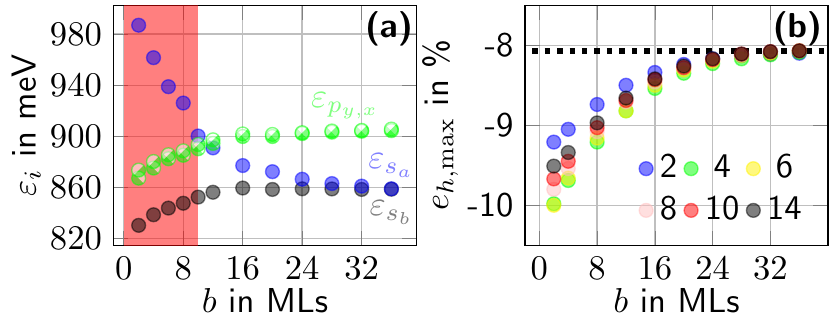}
	\caption{Evolution of the states absolute energy and hydrostatic strain. (a) Energetic position of Kramers-degenerate electronic states $\varepsilon_i$ as a function of the separating barrier width $b$ in a pair of two In$_{1.0}$Ga$_{0.0}$As QDs. Filled and semi-filled circles denote the binding $s_b$-, anti-binding $s_a$- and $p$-type orbitals (black, blue and green), respectively. (b) Maximum negative hydrostatic strain $e_{h,max}$ as a function of the barrier width $b$ for an isolated pair of coupled QDs (blue) and the two central QDs in stacks of 4 to 14. The dotted line represents the hydrostatic strain within a single QD.
	}
	\label{fig:fig_7}%
\end{figure}%
We compare the results of LCQO and full 8-band $\dotkp$ calculations for stacks of two identical In$_{x}$Ga$_{1-x}$As QDs.
Fig.\ \ref{fig:fig_4} shows the evolution of the two intra-band transitions $\ket{\psi_{p_y,_x}} \rightarrow \ket{\psi_{s_b}}$ 
in the system as a function of QD separation (coupling strength) and material composition for both methods.
The eigenstates $\ket{\psi_{p_y,_x}}$ and $\ket{\psi_{s_b}}$ have an odd ($p$-like) and even ($s$-like) symmetry \cite{schliwa_impact_2007}, respectively, allowing optical transitions in agreement with parity selection rules.
The results for the LCQO method and full 8-band $\dotkp$ calculations are indicated by blue circles and red squares, respectively.
In Fig.\ \ref{fig:fig_4}(a) the separating barrier width $b$ is varied between 2 and 36\,MLs in steps of 2\,MLs, for two In$_{1.0}$Ga$_{0.0}$As QDs, mapping the transition from strong to vanishing inter-dot coupling.
Fig.\ \ref{fig:fig_4}(b) shows the influence of an increasing gallium content up to 50\,\% (in steps of 10\,\%) on the transition energies at a constant barrier width of 16\,MLs.
In both series, the methods agree well in general trend and produce similar transition energies with deviations $\lesssim 4$\,meV, smaller than the overall tuning range.
The physical origin of the global minimum around $b=10$\,MLs in Fig.\ \ref{fig:fig_4}(a) will be discussed in detail in section III.C.
Fig.\ \ref{fig:fig_5} shows the probability density $|\psi_{i}|^{2}$ for the first four Kramers-degenerate electron states for $b=16$\,MLs as calculated using the full 8-band $\dotkp$ (top) and LCQO method (bottom).
Both probability density and symmetry of the eigenstates found using the LCQO method are in excellent agreement to the full 8-band $\dotkp$ results.
Overall, the LCQO method maps the electronic structure of coupled QDs excellently and provides realistic results comparable to a full $\dotkp$ calculation.
\subsection{Impact of hydrostatic strain and piezoelectricity on stacks of electronically coupled QDs} \label{cp:strain}
The lattice mismatch of $0.4\,$\AA\ between InAs and GaAs results in a highly strained heterostructure.
In a simplified picture the electronic states of the coupled QDs shift linearly with the hydrostatic strain $e_{h}$ \cite{chuang_physics_2009}.
Fig.\ \ref{fig:fig_6}(a) depicts the hydrostatic strain distribution for three exemplary systems of In$_{1.0}$Ga$_{0.0}$As QDs: a single QD and stacks of two and ten QDs, with a separating barrier width of $b=8$\,MLs.
The larger lattice constant of InAs leads to a compressive strain within the single QD in Fig.\ \ref{fig:fig_6}(a).
Subsequent stacking of QDs results in a cumulative compressive strain for the QDs in the center, while strain relaxation takes place mainly at the top and bottom of the stack. 
For the stack of ten QDs, the strain distribution is more or less uniform in the central region of the stack, cf.\ shaded area in Fig.\ \ref{fig:fig_6}.
This is also reflected in the distribution of the piezoelectric potential $\phi$ in Fig.\ \ref{fig:fig_6}(c). 
For a single QD, the piezoelectric potential vanishes inside the QD and shows a quadrupole-like distribution outside the QD, with inverted polarity at the top and bottom interfaces. 
As a result, the superimposed single QD potentials cancel each other out in the central region of the stack and are amplified at both ends of the chain, cf.\ Fig.\ \ref{fig:fig_6}(c). 
Fundamentally, the vanishing piezoelectric field inside the QDs and the symmetry of the quadrupole-like potential are linked to the first- and second-order piezoelectricity, as shown in \cite{schliwa_impact_2007}.
Both the inhomogeneous strain distribution and the resulting piezoelectricity impact the electronic structure by shifting the conduction band edge $V_{CB}$, cf.\ solid black line in Fig.\ \ref{fig:fig_6}(b).
In the uniform central region we expect to find cascades of electron states, delocalized over neighbouring QDs, suitable to generate the staircase potential required for a QCL active region.
The probability densities $|\psi_{i}(z)|^{2}$ for the three lowest Kramers-degenerate electron states are shown in Fig.\ \ref{fig:fig_6}(b) exemplarily.
As expected, the density of the first two states is localized within the QDs at the edges of the QD-chain having the lowest $V_{CB}$ potential. The third state's density (red) is already delocalized within the QD-chains central region and provides an $s$-type orbital constituting a 'ground state' within a staircase-potential, build-up at a certain applied external bias.
The energetic sequence of the orbitals, as they occur in coupled QDs, depends decisively on the coupling strength and thus on the inter-dot distances.
To illustrate this, we investigate the orbital symmetry of the lowest conduction band states in two coupled QDs as a function of the barrier width.
Fig.\ \ref{fig:fig_7}(a) shows the evolution of the energetic position $\varepsilon_i$ of the lowest Kramers-degenerate electron states for the stack of two QDs discussed in Fig.\ \ref{fig:fig_4}(a). 
For strong QD coupling, cf.\ shaded area in Fig.\ \ref{fig:fig_7}(a), we find the conventional single-QD $s_b$-/$p$-/$p$-type orbital sequence, with the binding $s$-type orbital as ground state, providing the two optically active intra-band transitions investigated throughout this work.
With increasing separation of the QDs, the anti-binding $s$-type orbital is lowered in energy, until for $b\gtrsim 12$\,MLs the orbital sequence is changed to $s_{b}$-/$s_{a}$-/$p$-/$p$-type.
Above a certain separation ($b=36$\,MLs for the QDs considered) the two $s$-type orbitals are no longer hybridized and can each be associated with one of the electronically uncoupled QDs.
A remaining energy difference is due to different strain states and long-range piezoelectric fields in both QDs.
Fig.\ \ref{fig:fig_7}(b) compares the hydrostatic strain state of an isolated system of two coupled QDs with the two QDs in the center of stacks of 4 to 14 QDs.
A clear difference is visible in the strong coupling regime.
The hydrostatic strain shows a maximum for the two QDs in the central region of the stacks of four and six QDs and then decreases again with number of QDs building the QD-chain. This is due to the lattice being deformed irregularly at the edges of the QD-chain, where the relaxation of the hydrostatic strain occurs.
In all cases, the hydrostatic strain state of a single QD is recovered for separations $b> 36$\,MLs.
\subsection{Intra-band transition energies as a function of QD size, material composition and coupling strength} \label{ch:barrier}
\begin{figure}[!t]
	\centering
	\includegraphics[]{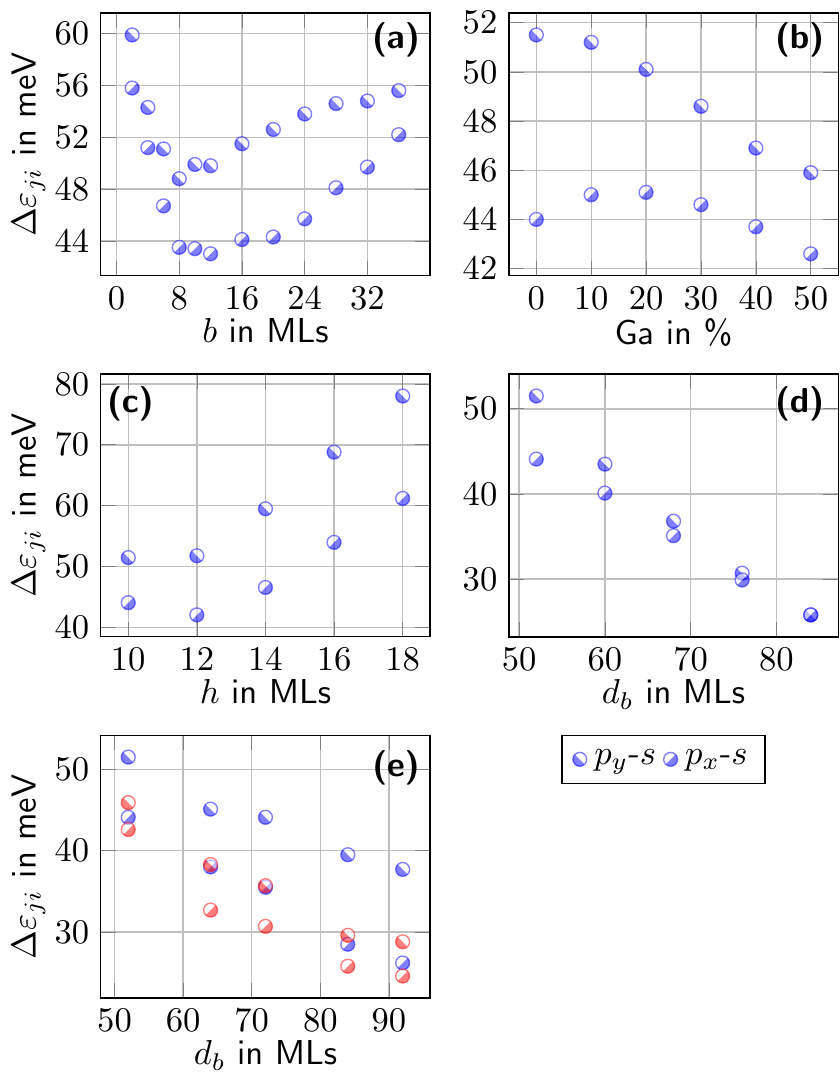}
	\caption{Evolution of the transition energy $\Delta \varepsilon_{ji}$ of the $p$-to-$s$ intra-band transitions for the two central QDs in a stack of ten identical QDs as a function of (a) barrier width $b$, (b) QD material composition and (c-e) size for a constant base diameter $d_{b}$, height $h$ and aspect ratio $AR_{v}$, respectively. Except for (b) and red symbols in (e), In$_{1.0}$Ga$_{0.0}$As/GaAs QDs are considered, separated by $b=16\,$MLs. The red semi-filled circles in (e) show the transition energies for In$_{0.5}$Ga$_{0.5}$As/GaAs QDs at a constant aspect ratio.
	}
	\label{fig:fig_8}%
\end{figure}
\begin{figure}[!t]
	\centering
	\includegraphics[]{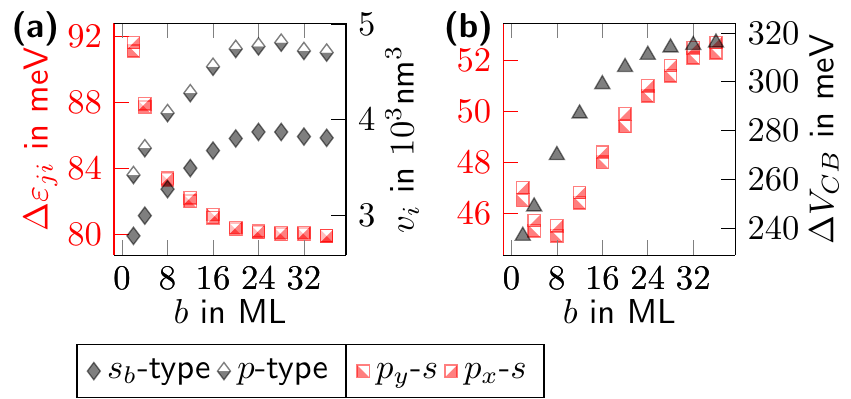} 
	\caption{Evolution of the energies $\Delta \varepsilon_{ji}$ of the $p$-to-$s$ intra-band transitions in stack of two identical QDs as a function of barrier width $b$ (red semi-filled squares). (a) The transition energies omitting strain as well as piezoelectricity and the volume $v_i$ of the corresponding binding $s$-type and the $p$-type orbitals (black filled and semi-filled diamonds). $v_i$ is the volume of the orbital $i$ at which the electron density is $\geq 0.1\%$ of its maximum value. 
	(b) The transition energies including strain but neglecting piezoelectricity. $V_{CB}$ depicts the depth of the QDs potential well measured in the mid of the top QD, i.e. the minimum of the conduction band edge. The transition energies where calculated using the 8-band $\dotkp$-model.
	}
	\label{fig:fig_9}%
\end{figure}%
In this section, we explore the influence of inter-dot coupling strength, QD material composition and QD size on the intra-band transition energies for a stack of ten model QDs.
Based on the knowledge from the preceding section, we investigate the electronic properties of the homogeneous central region of the QD stack, that is eligible for a staircase potential with constant energy spacings.
In this regard, ten is the minimum number of coupled QDs for the edge effects to be converged, leading to a flat conduction band edge and constant hydrostatic strain within the two central QDs, cf.\ shaded area in Fig.\ \ref{fig:fig_6}.
In Fig.\ \ref{fig:fig_8} the evolution of the $p$-to-$s$ intra-band transition energies in the central QD pair is shown as a function of separating barrier width $b$ (a) and QD material composition (b), as well as QD size (c-e), where height, base diameter and vertical aspect ratio are kept constant, respectively, for $b=16$\,MLs.
For strongly coupled QDs, we observe a red-shift of the transition energies with increasing barrier width, up to a global minimum at $b\approx 8$\,MLs, that coincides with the change in orbital sequence discussed above.
With increasing gallium content, the transition energy shows a maximum at In$_{0.8}$Ga$_{0.2}$As for the $p_x$-to-$s$ and a monotonic decrease for the $p_y$-to-$s$ transition. 
This is due to a $V_{CB}$ conduction band potential in the QDs increasing with the gallium content and also leading to a reduced energy splitting of the $p$-type orbitals.
Figs.\ \ref{fig:fig_8}(c) and (d) show a monotonic blue- and red-shift for an increasing QD height for a constant base diameter (i.\,e.\ an increasing aspect ratio) and a decreasing aspect ratio at constant height, respectively.
An energetic red-shift can achieved by increasing QD size at constant aspect ratio (e), with the red symbols showing In$_{0.5}$Ga$_{0.5}$As QDs, shifting the transitions further to lower energies and decreasing the splitting of the $p$-type orbitals, cf.\ Fig.\ \ref{fig:fig_8}(b).
We discuss the global minimum in Figs.\ \ref{fig:fig_8}(a) and \ref{fig:fig_4}(a) by considering the impact of barrier width and hydrostatic strain on the energy of the intra-band transitions.
Figs.\ \ref{fig:fig_9}(a) and (b) depict the energies of the $p$-to-$s$ transitions as a function of barrier width for a stack of two QDs, where in (a) strain as well as piezoelectric fields are neglected and in (b) strain is included but piezoelectricity is omitted.
With neglected strain and piezoelectricity, the transition energies show a monotonous decrease with increasing barrier width (red-shift).
This is linked to the increasing volume of the orbitals, cf.\ black diamonds in Fig.\ \ref{fig:fig_9}(a), since with increasing separation of the QDs, a growing part of the electron densities is located in-between the QDs. 
The density's volume decreases again for $b\geq 32$\,MLs, as the QDs decouple and the electrons more localized.
The inclusion of strain into the calculation of the transition energies shows already the global minimum, cf.\ Fig.\ \ref{fig:fig_9}(b). 
With increasing barrier width the depth of the QDs potential well $\Delta V_{CB}$, cf.\ black triangles in Fig.\ \ref{fig:fig_9}(b), is increasing as the hydrostatic strain decreases, cf. Fig.\ \ref{fig:fig_7}(b), resulting in a blue-shift of the transition energies. 
This effect dominates for barrier widths of $b \gtrsim 8$\,MLs, whereas in the strong coupling regime the red-shift resulting from an increasing volume of the electron orbitals is prevailing.
\subsection{Intra-band staircase potential design} \label{cp:bandstructure}
\begin{figure}[!t]
	\centering
	\includegraphics[]{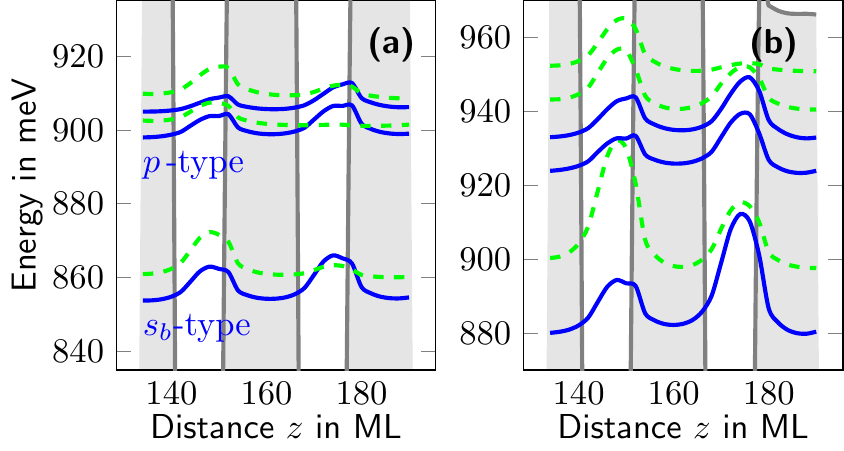}\\
	\medskip 
	\includegraphics[]{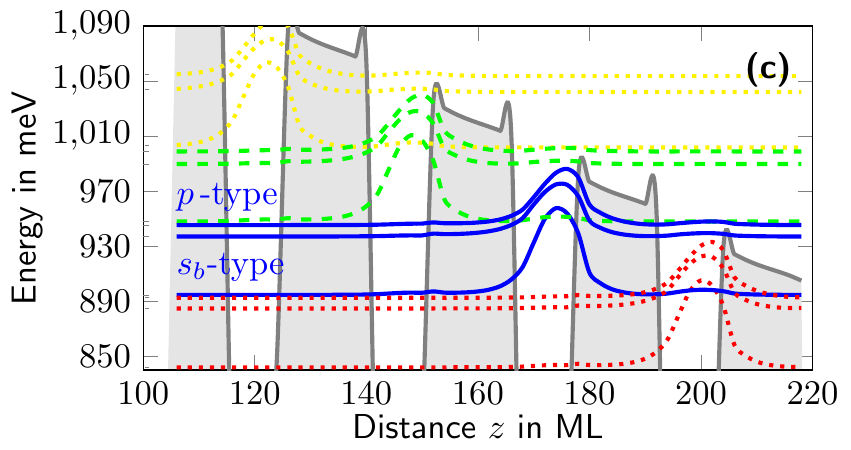}
	\caption{The $V_{CB}$ bandstructure staircase potential for the two central QDs in a stack of ten identical In$_{1.0}$Ga$_{0.0}$As/GaAs
	QDs separated by $b=16\,$MLs at various external biases $E_{\textrm{ext}}$. Gray lines and shaded areas depict the calculated conduction band-edge and barrier material, respectively. Colored lines show the $|\psi_{i}(z)|^{2}$ of the $s$- and $p$-type Kramers-degenerate electron states for the QDs in the central region showing a maximum of probability density. (a) The staircase potential without an external bias. (b) The staircase potential at $E_{\textrm{ext}}=12$\,kVcm$^{-1}$. Solid blue and dashed green lines depict the $s_b$- and $p$-type orbitals discussed in section \ref{ch:barrier}. (c) the staircase potential at $E_{\textrm{ext}}=36$\,kVcm$^{-1}$ showing four central QDs.}
	\label{fig:fig_10}%
\end{figure}%
\begin{figure}[!t]
	\centering
	\includegraphics[]{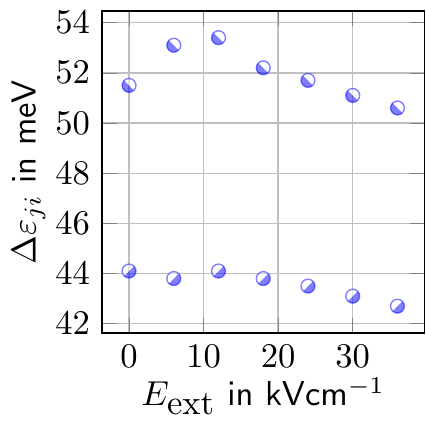}
	\caption{Evolution of the energy $\Delta \varepsilon_{ji}$ of the $p$-to-$s$ intra-band transitions for the two central QDs in a stack of ten In$_{1.0}$Ga$_{0.0}$As/GaAs QDs separated by barriers of $b=16\,$MLs as a function of external bias $E_{\textrm{ext}}$.
	}
	\label{fig:fig_11}%
\end{figure}%
\begin{figure}[!t]
	\centering
	\includegraphics[]{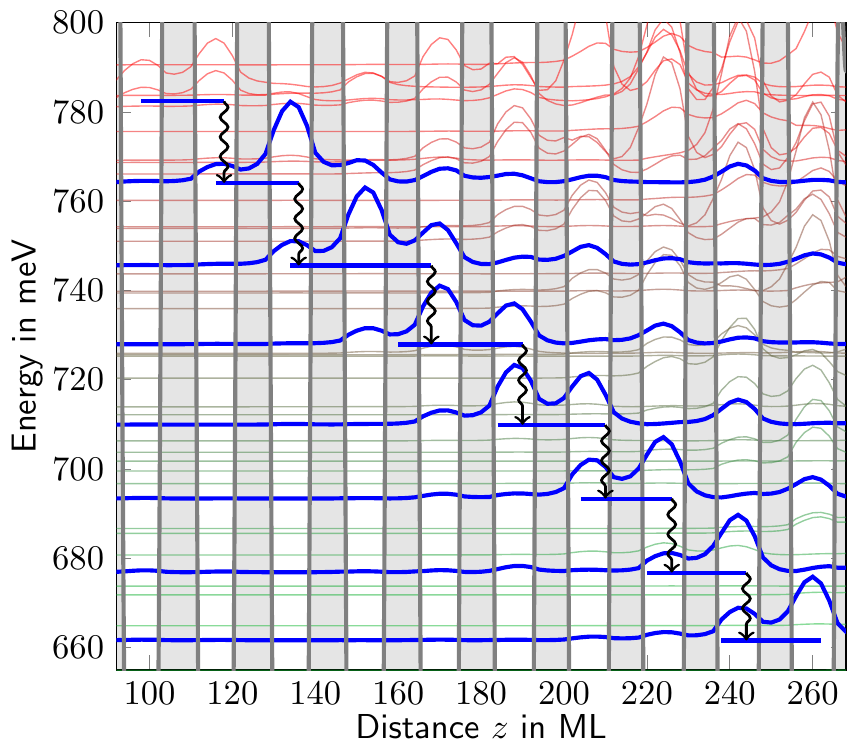}
	\caption{The $V_{CB}$ bandstructure staircase potential for a stack of 20 identical In$_{0.8}$Ga$_{0.2}$As/GaAs QDs with basis lengths and heights of $14.7\,$nm and $2.8\,$nm, respectively, at an external bias of $E_{ext}=16\,$kVcm$^{-1}$. The barrier width is set to $8\,$MLs. Wavy arrows indicate possible intra-band transitions. A plot of the full bandstructure staircase potential is provided in Supplemental Material at [URL].}
	\label{fig:fig_12}%
\end{figure}%
In this section, we show that an intra-band staircase potential of subsequent, identical intra-band transitions can be realized in such stacks of QDs.
In Fig.\ \ref{fig:fig_10} the conduction band staircase potential for the two central QDs in a stack of ten QDs at various external biases are shown illustrating its influence on the states energy and delocalization. 
Colored lines show the probability densities of the Kramers-degenerate electron states exhibiting a maximum within the two central QDs, whose $p$-to-$s$ transition energies correspond to the discussion in the previous section \ref{ch:barrier} and Fig.\ \ref{fig:fig_8}.
Figs.\ \ref{fig:fig_10}(a) and (b) show the $V_{CB}$ bandstructure without and with an external bias of $E_{\textrm{ext}}=12$\,kVcm$^{-1}$, respectively, which has basically three effects on the staircase potential: 
With increasing external bias, the energetic distances of the $s$- and $p$-type orbitals between the adjacent QDs (i.e. inter-dot $s$-to-$s$ or $p$-to-$p$ transition), depicted by solid blue and dashed green lines, respectively, are increasing.
In contrast, the energies of the $p$-to-$s$ intra-dot transitions remain almost constant, see Fig.\ \ref{fig:fig_11} for an evolution of the intra-dot transition energies as a function of external bias.
A third effect is the increasing localization of the states with the bias. Fig.\ \ref{fig:fig_10}(c) shows the staircase potential at an external bias of $E_{\textrm{ext}}=36$\,kVcm$^{-1}$, where the probability densities of the $s$- and $p$-type orbitals are almost completely located in their respective QDs.\\
Lastly, we examine the electronic structure of an exemplary stack of 20 coupled QDs under external bias and show that transition energies can be engineered to provide optical gain in the infrared spectral range.
Fig.\ \ref{fig:fig_12} shows a zoom into the relevant region of the conduction band of the stack of 20 identical In$_{0.8}$Ga$_{0.2}$As/GaAs QDs with a height of $5$\,MLs and a diameter of $36$\,MLs as discussed in section \ref{ch:results}, separated by $b=8$\,MLs, at an external bias of $E_{ext}=16\,$kVcm$^{-1}$. 
For the chosen number of QDs, at least seven transitions at an energy of $\approx 17.2$\,meV are present in the central part of the stack. After edge effects are converged, suitable transitions scale with the QD-chain length, i.e. the gain.
Carriers occupy, subsequently, the ground state within a QD and an excited state of the adjacent QD, building pairwise electronically coupled QDs along the chain. 
The orbitals show an $s$-type symmetry within a QD.
These transitions within coupled QDs, like the $p$-to-$s$ transitions in Fig.\ \ref{fig:fig_8}, can be tuned using barrier width, material composition and external bias allowing, for example, an effective relaxation via LO-phonons. 
To achieve population inversion, however, an additional $p$-to-$s$ or $s$-to-$p$ transition within the staircase potential is required. 
Due to the comparatively low intra-dot transition energies close to the materials LO-phonon energy, cf. Fig.\ \ref{fig:fig_8}, this transition could be achieved rather than a diagonal transition across an additional barrier. 
A suitable QCL bandstructure design based on coupled two-QD unit cells, enabling population inversion as well as providing a wavelength within the far-infrared, is developed and discussed in (the associated manuscript) Ref.\ [PRL]. 
\section{Conclusion}
We developed a novel method for the calculation of excited states in electronically coupled QD systems based on QD single-particle wave functions, which enables and accelerates the calculation of the bandstructure of QD stacks consisting of 20 and more QDs, including tens of electronic states at very low computational cost. Facilitated by the LCQO method, we have investigated the evolution of conduction band states as a function of various QD parameters, which is a prerequisite for the development of suitable active regions of QD-QCLs. With this we demonstrated an exemplary staircase potential with equally distributed energy transitions and probability densities at an external bias. Our calculations pave the way for the development of an active region of a QCL based on a QD superlattice, exploiting the intrinsic advantages of QDs, which can lead to low threshold current densities and elevated operating temperatures, in particular with far-infrared cascade lasers.
\begin{acknowledgments}
This work was in part funded by the Deutsche Forschungsgemeinschaft in framework of the SFB 787.
\end{acknowledgments}%


\bibliography{lib_korr}

\end{document}